\documentclass[conference]{IEEEtran}

\IEEEoverridecommandlockouts
\usepackage{cite}
\usepackage{amsmath,amssymb,amsfonts}
\usepackage{algorithmic}
\usepackage{graphicx}
\usepackage{textcomp}
\usepackage{xcolor}
\usepackage{float} 
\usepackage{orcidlink}
 
\def\BibTeX{{\rm B\kern-.05em{\sc i\kern-.025em b}\kern-.08em
    T\kern-.1667em\lower.7ex\hbox{E}\kern-.125emX}}
\usepackage{fancyhdr}
\usepackage{lastpage}
\usepackage[english]{babel}
\usepackage{tikz}
\usetikzlibrary{calc}
\usepackage{lipsum}

\usepackage{ amssymb }
\usepackage{ gensymb }
\usepackage{amsfonts}

\usepackage{mathtools}
\usepackage{amsmath}
\usepackage{tabularx}
\usepackage{anyfontsize}
\usepackage{multicol}

\usepackage{lettrine}
\begin{document}

\title{Lepton Flavour Violation Identification in Tau Decay ($\tau^{-} \rightarrow \mu^{-}\mu^{-}\mu^{+}$) Using Artificial Intelligence \\
}

\author{\IEEEauthorblockN{Reymond Mesuga}
\IEEEauthorblockA{\textit{Department of Physical Sciences, College of Science} \\
\textit{Polytechnic University of the Philippines, Manila, Philippines} \\
\text{Corresponding author: rrmesuga@iskolarngbayan.pup.edu.ph}}

}
\maketitle
\begin{abstract}
The discovery of neutrino oscillation, proving that neutrinos do have masses, reveals the misfits of particles in the current Standard Model (SM) theory. In theory, neutrinos having masses could result in lepton flavour not being a symmetry called Lepton Flavour Violation (LFV). While SM theory extensions allowed LFV processes, their branching fractions are too small, making them unobservable even with the strongest equipment up-to-date. With that, scientists in recent years have generated LFV-like processes from the combined LHCb and Monte-Carlo-Simulated data in an attempt to identify LFV using Artificial Intelligence (AI), specifically Machine Learning (ML) and Deep Learning (DL). In this paper, the performance of several algorithms in AI has been presented, such as XGBoost, LightGBM, custom 1-D Dense Block Neural Networks (DBNNs), and custom 1-D Convolutional Neural Networks (CNNs) in identifying LFV signals, specifically $\tau^{-} \rightarrow \mu^{-}\mu^{-}\mu^{+}$ decay from the combined LHCb and Monte-Carlo-Simulated data that imitates the signatures of the said decay. Kolmogorov-Smirnov (KS) and Cramer-von Mises (CvM) tests were also conducted to verify the validity of predictions for each of the trained algorithms. The result shows decent performances among algorithms, except for the LightGBM, for failing the CvM test, and a 20-layered CNN for having recorded a considerably low AUC. Meanwhile, XGBoost and a 10-layered DBNN recorded the highest AUC of 0.88. The main contribution of this paper is the extensive experiment involving custom DBNN and CNN algorithms in different layers, all of which have been rarely used in the past years in identifying LFV-like signatures, unlike GBMs and tree-based algorithms, which have been more popular in the said task.
\end{abstract}


\section{\textbf{Introduction}}
\subsection{\textbf{The Standard Model}}
\begin{figure}[t]
\includegraphics[width=0.9999\columnwidth]{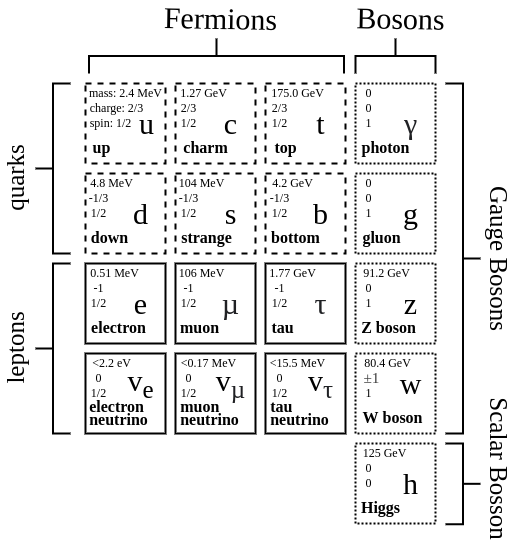}
\caption{The Standard Model (SM)}
\label{fig:1}
\end{figure}

\noindent
\lettrine[findent=0pt]{\textbf{I}}{ }n particle physics, the Standard Model (SM) is a theory that describes all known elementary particles, also known as fermions (see Fig. \ref{fig:1}), as well as three of nature's four fundamental forces (i.e., electromagnetic, weak nuclear, and strong nuclear force). It is the theory that is closest to the "Theory of Everything" as it summarizes all the fundamental particles and forces that we are currently aware of as of this age. It is even called "The Theory of Almost Everything" by professor Robert Oerter in his famous book that describes the beauty and imperfection of SM \cite{Oerter}. As shown in Fig. \ref{fig:1}, elementary particles are classified into two types: fermions and bosons. Fermions are particles that follow the rules of Fermi-Dirac statistics, which is a type of quantum statistics that deals with the physics of a system made up of many identical particles that follow Pauli's exclusion principle, which states that in a quantum system, two or more identical particles with half-integer spins (i.e. fermions) cannot occupy the same quantum state at the same time \cite{Saunders}. Bosons, on the other hand, are fundamental forces that mediate the fundamental interactions of strong, weak, and electromagnetic forces. There are two types of fermions, namely quarks and leptons. Quarks are the fundamental building blocks of matter and can be found inside protons and neutrons where the quarks bind together through the strong nuclear force to form such particles. Meanwhile, leptons only interact through electromagnetic and weak nuclear forces. Charged lepton flavours are electrons, muons, and taus. All of the charged lepton flavours have their own neutrino, i.e., electron-neutrino, muon-neutrino, and tau-neutrino, respectively. SM provides the most up-to-date explanation of how particles interact with each other at a quantum scale, with very high accuracy, backed up by several successful experiments conducted in different particle accelerators. The latest being the discovery of the Higgs boson in 2012, which was first introduced in the SM in 1964. Despite its success, SM is still lacking an explanation when it comes to neutrino oscillations, dark matter, and the concept of gravity on a quantum scale.

\subsection{\textbf{Lepton Flavour Violation}}

One of the biggest challenges in the field of particle physics up-to-date is the detection of rare phenomena that could violate the current rules of SM. In SM, it is assumed that neutrinos don't have masses. However, neutrino oscillation experiments implies that neutrinos do have non-zero masses which results to the modification of the SM. Given a measurement, the probability of finding a given flavour is:

\begin{equation}
\label{eq:1}
P(v_{e}-v_{i}) \approx U_{ei}^{2}sin^{2}(\frac{\Delta m^{2}L}{4E} )
\end{equation}

which also implies neutrinos having mix angles resulting to lepton flavor not being a symmetry. In nature, some physical quantities  are conserved such as energy and momentum. According to Noether's Theorem, both conservation law is associated with a fundamental symmetry which is crucial to the structure of SM \cite{Leone}. Since, neutrinos where suggest to have non-zero mass according to neutrino oscillation experiment, possible extensions in SM suggests that leptons are not a symmetry in nature which could lead to a phenomena that could violate lepton favour decay famously called Lepton Flavour Violation (LFV). LFV processes are allowed in SM but their branching fractions are of order $10^{-40}$ which makes them almost impossible to observe even by the strongest colliders up-to-date \cite{Raidal}. 

The type of LFV decay to be identified in this work will be the $\tau$ decay, specifically, $\tau \rightarrow \mu\mu\mu$. At $e^{+}e^{-}$ colliders, researchers have looked for LFV specifically in $\tau$ decays. The branching ratio $B(\tau \rightarrow \mu\mu\mu)$ has been subjected to the most strict limits specified by the B factories, i.e., BaBar and Belle at the order of $10^{-8}-10^{-7}$ \cite{Amhis}. In addition, Belle collaboration reports $B<2.1\times10^{-8}$ for $\tau \rightarrow \mu\mu\mu$ decay \cite{Hayasaka}. The objective of this work is to identify $\tau \rightarrow \mu\mu\mu$ decay using the mix of real data from LHCb experiment and monte-carlo simulated dataset decay both of which will be used to train artificial intelligence algorithm which will be discussed in the next sections.

\subsection{\textbf{Artificial Intelligence}}
In this work, the term Artificial Intelligence (AI) is used to describe both Machine Learning (ML) and Deep Learning (DL)) algorithms as they are both sub-fields of AI. AI has been successful in the last decade for its application in many fields such as healthcare, military security, the stock market, automobiles, etc. Just recently, AI was already introduced in the field of physical sciences as well, such as gravitational physics in identifying glitch wave-forms, classical mechanics in answering three-body problems, and especially particle physics in identifying exotic particles. The following subsections will discuss two of the most popular algorithms in AI, i.e., Gradient Boosting Machines in ML and Artificial Neural Networks in DL.

\subsubsection{\textbf{Gradient Boosting Machine (GBM)}}The boosting algorithm, which is a process of combining several reasonably accurate predictions to create one exceedingly accurate prediction, was proposed by \cite{Freund}. GBM, in general, uses back-fitting and non-parametric regressions to create prediction models. Rather than creating a single model, the GBM creates an initial model and then fits new models using loss function minimization to produce the most accurate model possible \cite{Friedman, Natekin}. There are many GBM algorithms that have already been introduced in the literature, such as AdaBoost, Catboost, LightGBM, XGBoost, etc. The GBM algorithm to be used in this work is XGBoost. The prediction scores of each decision tree algorithms in a GBM algorithm can be written mathematically as

\begin{equation}
\label{eq:2}
\hat{y}_{i} = \sum_{k=1}^{K}f_{k}(x_i), f_{k} \in \mathcal{F}
\end{equation}

 \noindent where $f$ is the function from the set of possible classification and regression trees (CARTs) symbolized as $\mathcal{F}$. The objective function for the above classifier can be written as

\begin{equation}
\label{eq:3}
obj(\theta) = \sum_{i}^{n}l(y_{i},\hat{y}_{i}) + \sum_{k=1}^{K} \Omega(f_{k})
\end{equation}

The above equation is the combination of loss function (first term) and the regularization parameter (second term). Rather than learning the tree all at once, which makes optimization more difficult, additive strategy is used, minimize the loss of previous trees, then add a new tree. Also, to employ the above objective function (i.e., eq. \ref{eq:3}), the said function must convert to a Euclidean domain function. To do that, Taylor's theorem is applied to eq. \ref{eq:3}, specifically, second-order approximation such that 

\begin{equation}
\label{eq:4}
\begin{gathered}
\hat{y}_{0}=0\\
\hat{y}_{i}^{(1)} = f_{1}(x_{i}) = \hat{y}_{i}^{(0)} + f_{i}(x_{i})\\
\hat{y}_{i}^{(2)} = f_{i}(x_{i}) + f_{2}(x_{i}) = \hat{y}_{i}^{(1)} + f_{2}(x_{i})\\
\dotsb\\
\hat{y}_{i}^{(t)} = \sum_{k=1}^{(t)}f_{k}(x_{i})=\hat{y}_{i}^{(t-1)} + f_{t}(x_{i}) 
\end{gathered}
\end{equation}

\noindent After applying the second-order approximation, the objective function will be

\begin{equation}
\label{eq:5}
\begin{gathered}
obj^{(t)} = \sum_{i=1}^{n}[l(y_{i},y_{i}^{(t-1)})+g_{i}f_{t}(x_{i})+\\\frac{1}{2}h_{2}f_{t}^{2}(x_{i})]+\Omega(f_{t}) + C,\;\; \text{where,} \\
\end{gathered}
\end{equation}

\begin{gather}
\label{eq:6}
g_{i} = \partial_{\hat{y}^{(t-1)}}l(y_{i},\hat{h}_{i}^{(t-1)} )\\
\label{eq:7}
h_{i} = \partial_{\hat{y}_{i}^{(t-1)}}^{2}l(y_{i},\hat{y}_{i}^{(t-1)})\\
\label{eq:8}
\Omega(f) = \gamma T + \frac{1}{2}\lambda\sum_{j=1}^{T} w_{j}^{2}
\end{gather}

 \noindent Simplifying eq. \ref{eq:5} results to the following:

\begin{gather}
\label{eq:9}
obj^{(t)} = \sum_{j=1}^{T}[G_{j}w_{j}+\frac{1}{2}(H_{j}+\lambda)w_{j}^{2}]+\gamma T, \;\; \text{where,} \\
G_{j} = \sum_{i\in I_{j}}g_{i} \;\; \text{and} \;\; H_{j}=\sum_{i\in I_{j}}h_{i}
\end{gather}

 \noindent The goal of a GBM algorithm is to find the most optimized output value for the leaf to minimize the whole equation (i.e., eq. \ref{eq:9}). This can be achieved by adding more decision trees that are better than previous trees.
 
 \begin{figure}[!h]
\includegraphics[width=0.9999\columnwidth]{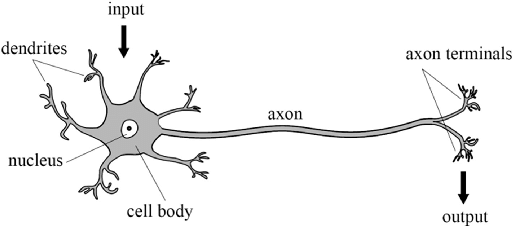}
\caption{Biological Neuron, \cite{Neves}}
\label{fig:2}
\end{figure}

\subsubsection{\textbf{Artificial Neural Network}}
\begin{figure}[!h]
\includegraphics[width=0.9999\columnwidth]{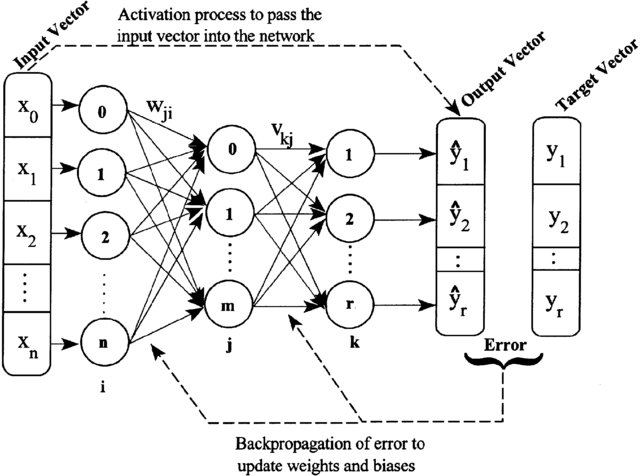}
\caption{Artificial Neural Network (ANN), \cite{Basheer}}
\label{fig:3}
\end{figure}

Artificial Neural Network (ANN) based algorithms have been among the most successful algorithms in the field of AI. It imitates the mechanism of the human brain and, at the same time, solves more complex problems by learning through the data. ANN has an artificial neuron, also called a perceptron, which is synonymous to the biological neuron shown in Fig. \ref{fig:2}. A biological neuron accepts input signals through its synapses located on the dendrites or the membrane of the neuron. If the input signal received by the synapses is strong enough to activate a certain threshold, it will then activate the neuron and release signal through the axon. The signal will probably be sent to other synapses, leading to the activation of further neurons. That mechanism of receiving an input signal and releasing an output is almost the same as in perceptrons. Perceptrons are the building blocks in ANN. The basic structure of an ANN can be visualized in Fig. \ref{fig:3}. It has three main parts, namely the input layer, the hidden layer, and the output layer. The input layer receives input signals (i.e., $X= [x_{0},x_{1},x_{2},x_{3},...,x_{n}]$) which is synonymous with the synapses of biological neurons. It will then apply randomly selected weights (i.e., $W = [w_{0},w_{1},w_{2},w_{3},...,w_{n}]$) on each input signal and a bias, such that

\begin{equation}
\begin{gathered}
\label{eq:11}
\hat{y} = XW + b  = x_{0}w_{0} + x_{1}w_{1}\\ + x_{2}w_{2} + x_{3}w_{3} + ... + x_{n}w_{n} + b\\
\hat{y} = \sum_{i=0}^{n}x_{i}w_{i} + b
\end{gathered}
\end{equation}

\noindent where $\hat{y}$ is the output of the perceptron, $x$ is an input signal, $w$ is the weight, and $b$ is the bias. Bias $b$ is some kind of special weight. While weight $w$ decides how fast is the activation function will activate, bias $b$ is used to delay the activation. The hidden layers will then receive the input signals passed by input layer and apply non-linear transformation through activation function. The most common activation used in hidden layers is the ReLU activation function defined as in equation \ref{eq:12}.

\begin{gather}
\label{eq:12}
f(x) = max(0,x)\\
\label{eq:13}
\hat{y} = \varphi(\sum_{i=0}^{n}x_{i}w_{i} + b)
\end{gather}

\noindent Equation \ref{eq:12} can be interpreted as any value less an zero will be automatically converted into 0. Meanwhile, $\varphi$ is the applied activation function on a perceptron output $y$. Since the set weight $w$ are randomly chosen during the receiving of input signals in the input layer, ANN would likely give a meaningless result.

ANN need to be trained in order to learn and give meaningful output. During training, the set of weights $W$ for every input signal will be adjusted as long as the loss has not reached its minimum value. In simple terms, loss can be defined as in eq. \ref{eq:14}.

\begin{gather}
\label{eq:14}
Loss = |(\hat{y}-y)| \\
\label{eq:15}
Loss = - \frac{1}{N}\sum_{i=1}^{N} y_{i}log \hat{y}_{i} + (1-y_{i})log(1-\hat{y}_{i})\\
\label{eq:16}
Loss = - \sum_{i=1}^{N} y_{i}log\hat{y}_{i}
\end{gather}

In more complicated terms, eq. \ref{eq:15} is the Binary Cross Entropy Loss which is commonly used to define $Loss$ for binary classification problems. For categorical classification (i.e., dataset that involves more than two classes), $Loss$ can be defined as in eq. \ref{eq:16} which is the Categorical Cross Entropy Loss. Specifically, $\hat{y}_{i}$ is the $i$-th scalar value in ANN output, $y_{i}$ is the target or actual value., and $N$ is the number of scalar values in the ANN output. Loss function is basically used as a reference to adjust the weights. The lower the loss, the better the performance of an ANN algorithm. The algorithm used to adjust the weights using $Loss$ is called the $Optimizer$. In basic terms, an optimizer is used to adjust the weights $W$ using $Loss$ can be defined as in eq. \ref{eq:17}.

\begin{gather}
\label{eq:17}
W_{new} = W_{old}-\alpha[\frac{\partial(Loss)}{\partial W_{old}}]
\end{gather}

\noindent where $W_{old}$ is the set of previous weights, $W_{new}$ is the new set of weights, $\alpha$ is the learning rate which controls how fast is the $Loss$ to reach its minimum. In literature, many advance optimizer algorithm is used such as Gradient Descent, Strochastic Grdient Descent, Adagrad, Adadelta, RMSprop, and Adam. The most up-to-date optimizer mentioned is Adam which is also used during the experiment of this work.

As a summary on how an ANN algorithm works, its goal is to imitate the human brain to automate and solve complicated task. It has three main parts namely the input layer, hidden layers, and out layer. Input layer receives and apply random weights on input signals. Weights can be interpreted as the "importance" of a signal. The weighted input signal will the passed to the hidden layers which will be applying the non-linear transformation of each inputs. The ANN will then gave a prediction which will be compared to the actual target value to compute the loss. The lower the loss, the better the performance of the model is when making predictions. The loss will then be used by the optimizer algorithm to adjust the weights. The adjustment of the weights will continue until the loss has reached its minimum value. That is the basic explanation on how ANN work and how it learn. 

\section{\textbf{Research Methods}}

\subsection{\textbf{Data Collection}}
The dataset used in this work were provided by \textit{Flavours of Physics: Finding $\tau \rightarrow \mu\mu\mu$} competition on Kaggle \cite{Kaggle}. The dataset is suited for binary-classification with the target labels of "signal" and "background." It contains both Monte-Carlo simulated data and real data from LHCb. The simulated ones are the signals, while the background ones were from LHCb. The features of the dataset have two main categories. First are the kinetic features, such as muon particle energy and momenta in a convenient coordinate system. Second are the isolation variables, as well as the impact parameters and numerous angles that characterize the particle's trajectory. The collected files from the competition are as follows:

\begin{table}[h]
\caption{Files in Dataset}
\label{tab:1}
\begin{tabularx}{\columnwidth}{ | l | X | }
\hline
    Filename & Description  \\ \hline
    \textbf{training.csv} & Data used in training algorithms. \\ \hline
    \textbf{check\_correlation.csv} & Used in performing Cramer\-von Mises (cvm) test on the trained algorithms.\\ \hline
    \textbf{check\_agreement.csv} & Used in performing Kolmogorov–Smirnov (KS) test and reporting the final performance in AUC of the trained algorithms. \\ \hline
    \textbf{test.csv} & Used in submitting the performance report in the competition. \\ \hline
\end{tabularx}
\end{table}

\subsection{\textbf{Classification Algorithms}}
Both deep learning and machine learning algorithms are implemented as follows:
\subsubsection{\textbf{Gradient Boosting Machines}} The two GBMs presented in this work are XGBoost and LightGBM. While two of the said GBMs have been experimented, only XGBoost did pass all the test required to be regarded as valid (dicussed further in section \ref{Evaluation}). The hyperparameters used during the training of XGBoost are as follows: eta=0.5, max\_depth=20, min\_child\_weight=3, subsample=0.5, and colsample\_bytree=0.7.

\subsubsection{\textbf{Artificial Neural Network Based Algorithms}}
ANN based algorithms were also implememted such as Dense Block Neural Networks (DBNN) and Convolutional Neural Networks (CNN) all of which are one dimensional. The parameters of the algorithms with respect to their number of layers are shown in Table \ref{tab:2}.

\begin{table}[ht]
\centering
\caption{No. of Parameters per Algorithm}
\label{tab:2}
\begin{tabular}{|l|l|l|l|} 
\hline
Algorithm      & Trainable Params                               & Non-trainable Params                       & Total Params                                    \\ 
\hline
5L-DBNN  & \textcolor[rgb]{0.129,0.129,0.129}{16,902,191} & \textcolor[rgb]{0.129,0.129,0.129}{20,480} & \textcolor[rgb]{0.129,0.129,0.129}{16,922,671}  \\ 
\hline
10L-DBNN & \textcolor[rgb]{0.129,0.129,0.129}{37,904,431} & \textcolor[rgb]{0.129,0.129,0.129}{40,960} & \textcolor[rgb]{0.129,0.129,0.129}{37,945,391}  \\ 
\hline
20L-DBNN & \textcolor[rgb]{0.129,0.129,0.129}{79,908,911} & \textcolor[rgb]{0.129,0.129,0.129}{81,920} & \textcolor[rgb]{0.129,0.129,0.129}{79,990,831}  \\ 
\hline
5L-CNN   & \textcolor[rgb]{0.129,0.129,0.129}{29,369,345} & 0                                          & \textcolor[rgb]{0.129,0.129,0.129}{29,369,345}  \\ 
\hline
10L-CNN  & \textcolor[rgb]{0.129,0.129,0.129}{31,475,713} & 0                                          & \textcolor[rgb]{0.129,0.129,0.129}{31,475,713}  \\
\hline
20L-CNN  & \textcolor[rgb]{0.129,0.129,0.129}{73,439,233} & 0                                          & \textcolor[rgb]{0.129,0.129,0.129}{73,439,233}  \\
\hline
\end{tabular}
\end{table}

\subsection{\textbf{Evaluation Tests and Metric}}
\label{Evaluation}
\subsubsection{\textbf{Kolmogorov-Smirnov (KS) Test}}
Since the classification algorithms are trained on both simulated and real-world data, it is feasible to get high performance by selecting features that aren't perfectly captured in the simulation. When applied to real and simulated data, the algorithms must not have a significant discrepancy. For that reason, a control channel, $Ds \rightarrow \varphi\pi$, is used which has a similar topology as the signal decay, $\tau \rightarrow 3\mu$. It's just that $Ds \ \varphi\pi$ is more popular, well-oberved behaviour, because it happens more frequently. The discrepancies between the classifier distributions on each sample are evaluated using the Kolmogorov–Smirnov (KS) test. All of the trained algorithms have to pass the KS test in order to be regarded as valid. The ideal KS value of a valid algorithms is below 0.09. Also, the agreement test set is used to execute KS test.
\subsubsection{\textbf{Cramer-von Mises (CvM) Test}}
Each particle is expected to have their own respected mass that differs from one another. In reality, mass is an estimate, and it is not a characteristic that scientists rely on when creating models. Correlations with mass can produce a false signal-like mass peak or erroneous background estimates. The test dataset does not have the mass column. A Cramer-von Mises (cvm) test, on the other hand, is conducted using hidden mass information.  If all of your mass sub-regions have similar distributions, your classifier is not correlated with the mass. CvM test value should be 0.002. Like the KS test, the trained algorithms were also required to pass the CvM test in order to be regarded to have valid predictions. 
\subsubsection{\textbf{Area Under the Curve (AUC) of ROC}}
\begin{figure}[!h]
\includegraphics[width=0.9999\columnwidth]{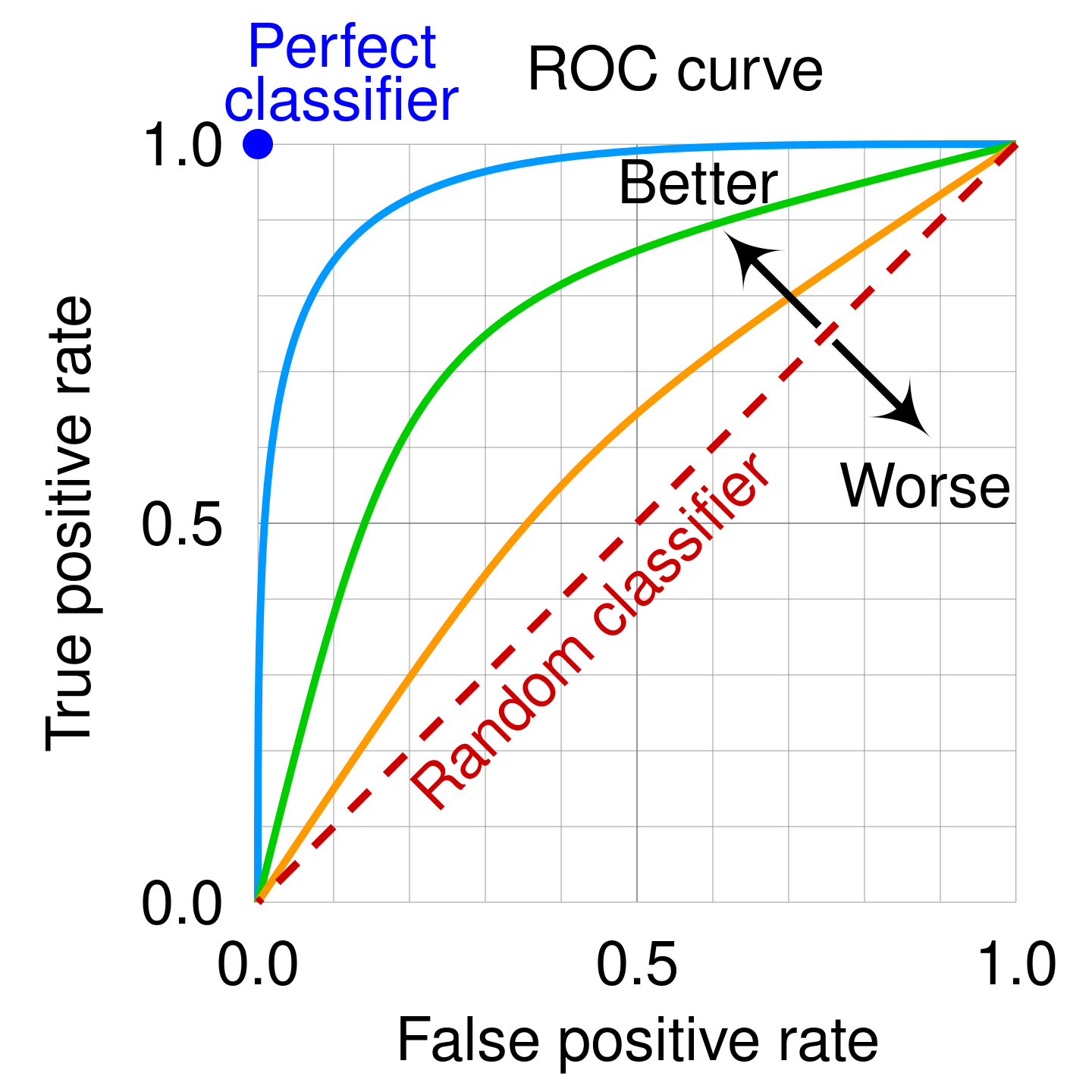}
\caption{Area Under the Curve of ROC, \cite{Wikipedia}}
\label{fig:4}
\end{figure}

AUC-ROC (or AUC) is commonly used in binary classification problems, but it also works in multi-class classification problems using one-vs-all precision-recall curves. AUC-ROC is primarily used to determine whether the classifier was successful in distinguishing between data classes. To plot the ROC, the True Positive Rate (TPR) and False Positive Rate (FPR) must be computed with several different thresholds and plotted as shown in Figure \ref{fig:4} where the x axis represents the FPR and the y axis represents the TPR. The optimal AUC score is 1, indicating that the classifier separated the classes in the data successfully. The lowest value is 0.5, indicating that the classifier makes predictions at random. The AUC metric was used as the primary performance metric in this work because it is the required performance metric for the competition.

\section{\textbf{Discussion of Results}}
Table \ref{tab:3} shows the KS and CvM test results that verify the validity of the algorithms. Despite failing at least one of the required tests, LightGBM is still included to present the results. Specifically, LightGBM failed the CvM test. The rest of the algorithms have passed both the KS and CvM tests, and thus can be regarded as algorithms with valid predictions.

\begin{table}[!h]
\centering
\caption{KS and CvM Test Results}
\label{tab:3}
\begin{tabular}{|l|l|l|l|} 
\hline
Algorithm     & KS Test                                           & CvM Test                                       & STATUS   \\ 
\hline
XGBoost  & 0.085993876                                       & 0.00090358                                     & PASSED   \\ 
\hline
5L-DBNN~ & \textcolor[rgb]{0.129,0.129,0.129}{0.054368892}   & \textcolor[rgb]{0.129,0.129,0.129}{0.00090938} & PASSED   \\ 
\hline
10L-DBNN & \textcolor[rgb]{0.129,0.129,0.129}{0.037683539}   & \textcolor[rgb]{0.129,0.129,0.129}{0.00110107} & PASSED   \\ 
\hline
20L-DBNN & \textcolor[rgb]{0.129,0.129,0.129}{0.043216691}   & \textcolor[rgb]{0.129,0.129,0.129}{0.00101158} & PASSED   \\ 
\hline
5L-CNN   & \textcolor[rgb]{0.129,0.129,0.129}{0.046371380}   & \textcolor[rgb]{0.129,0.129,0.129}{0.00104183} & PASSED   \\ 
\hline
10L-CNN  & \textcolor[rgb]{0.129,0.129,0.129}{0.067774241}   & \textcolor[rgb]{0.129,0.129,0.129}{0.00108337} & PASSED   \\ 
\hline
20L-CNN  & \textcolor[rgb]{0.129,0.129,0.129}{0.070314698}   & \textcolor[rgb]{0.129,0.129,0.129}{0.00116837} & PASSED   \\
\hline
LightGBM & \textcolor[rgb]{0.129,0.129,0.129}{5.5448286e-05} & \textcolor[rgb]{0.129,0.129,0.129}{0.06144420} & FAILED~  \\ 
\hline
\end{tabular}
\end{table}

\begin{figure}[!h]
\includegraphics[width=0.9999\columnwidth]{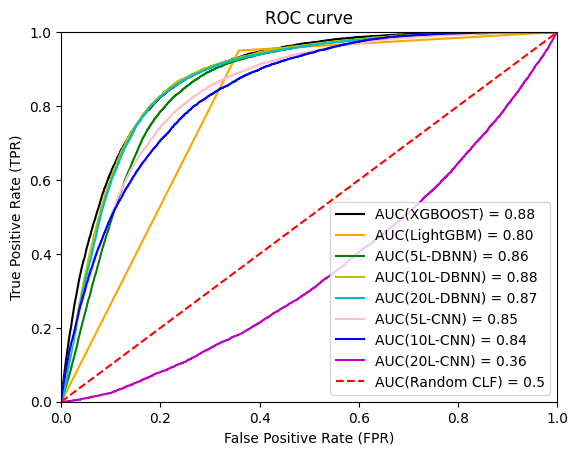}
\caption{Overall Result in AUC}
\label{fig:5}
\end{figure}

The AUC results for each algorithm indicate their capability to distinguish signals from the background. As illustrated in Fig. 5, XGBoost, a decision tree-based algorithm, is capable of identifying signals comparable to the performance of DL algorithms such as DBNN and CNN. In fact, the algorithms XGBoost and 10 layered DBNN (i.e., 10L-DBNN) have recorded the highest AUC among other experimented algorithms, with a value of 0.88 followed by a 20 layered DBNN (i.e., 20L-DBNN) and a 5 layered DBNN (i.e., 5L-DBNN) with an AUC of 0.87 and 0.86, respectively. LightGBM recorded a considerably decent AUC of 0.80 despite failing the CvM. Meanwhile, the 20 layered CNN (i.e., 20L-CNN) recorded the lowest AUC of 0.36, which is a lot worse than a random classifier, which is probably the surprising part of this result. Also, the 20L-CNN could possibly suffer from over-fitting, resulting in poor performance. Overall, the group of dense block neural networks (i.e., DBNN) has shown better performance than other groups, but XGBoost is competing as well.

\section{\textbf{Conclusion}}
While LFV processes are not yet directly observed due to the limitations of current technologies, experiments like neutrino oscillation suggest a very strong LFV is, indeed, a real thing. While LFV processes are allowed in the current SM theory, their branching fraction is too small, making them unobservable even with the strongest colliders up-to-date. With that, scientists attempted to use LHCb data and Monte Carlo simulated data to produce LFV-like features which can be used in AI applications. In this work, most of the AL algorithms trained to identify LFV using LHCb and Monte Carlo data are, indeed, capable of identifying LFV processes, specifically, $\tau \rightarrow \mu\mu\mu$ at decent AUC. Although, the competition focusing on this specific task has already been conducted, the results in this paper could help scientists in decision making, especially the results on the performance of several custom 1-D CNN and DBNN, all of which have been rarely used in the said competition. With that, the main contribution of this paper is the demonstration and reporting of the performance of several custom DBNN and CNN algorithms, all of which have been rarely used in the past years in identifying LFV, unlike GBMs and tree-based algorithms, which have been popular in the said task.

\section{\textbf{Author Statement}}
The writing of the manuscript and conducting of the whole experiment to present the results in this work were entirely done by a single author.




\bibliographystyle{IEEEtran}
\bibliography{conference_101719}

\begin{thebibliography}{10}
\providecommand{\url}[1]{#1}
\csname url@samestyle\endcsname
\providecommand{\newblock}{\relax}
\providecommand{\bibinfo}[2]{#2}
\providecommand{\BIBentrySTDinterwordspacing}{\spaceskip=0pt\relax}
\providecommand{\BIBentryALTinterwordstretchfactor}{4}
\providecommand{\BIBentryALTinterwordspacing}{\spaceskip=\fontdimen2\font plus
\BIBentryALTinterwordstretchfactor\fontdimen3\font minus
  \fontdimen4\font\relax}
\providecommand{\BIBforeignlanguage}[2]{{%
\expandafter\ifx\csname l@#1\endcsname\relax
\typeout{** WARNING: IEEEtran.bst: No hyphenation pattern has been}%
\typeout{** loaded for the language `#1'. Using the pattern for}%
\typeout{** the default language instead.}%
\else
\language=\csname l@#1\endcsname
\fi
#2}}
\providecommand{\BIBdecl}{\relax}
\BIBdecl

\bibitem{Oerter}
R.~Oerter and B.~Holstein, ``The theory of almost everything: The standard
  model, the unsung triumph of modern physics,'' \emph{Physics Today - PHYS
  TODAY}, vol.~59, 07 2006.

\bibitem{Saunders}
\BIBentryALTinterwordspacing
S.~Saunders, \emph{Fermi-Dirac Statistics}.\hskip 1em plus 0.5em minus
  0.4em\relax Berlin, Heidelberg: Springer Berlin Heidelberg, 2009, pp.
  230--235. [Online]. Available:
  \url{https://doi.org/10.1007/978-3-540-70626-7_71}
\BIBentrySTDinterwordspacing

\bibitem{Leone}
R.~Leone, ``On the wonderfulness of noether's theorems, 100 years later, and
  routh reduction,'' \emph{arXiv: History and Philosophy of Physics}, 2018.

\bibitem{Raidal}
M.~Raidal \emph{et~al.}, ``\BIBforeignlanguage{English (US)}{Flavor physics of
  leptons and dipole moments},'' \emph{\BIBforeignlanguage{English
  (US)}{European Physical Journal C}}, vol.~57, no. 1-2, pp. 13--18, Sep. 2008.

\bibitem{Amhis}
Y.~Amhis \emph{et~al.}, ``Averages of b-hadron, c-hadron, and $\tau
  \tau$-lepton properties as of summer 2016,'' \emph{The European Physical
  Journal C}, vol.~77, pp. 1--335, 2010.

\bibitem{Hayasaka}
K.~Hayasaka, K.~Inami, and Y.Miyazaki, ``Search for lepton flavor violating tau
  decays into three leptons with 719 million produced tau+tau- pairs,'' 2010.

\bibitem{Freund}
\BIBentryALTinterwordspacing
Y.~Freund and R.~E. Schapire, ``A decision-theoretic generalization of on-line
  learning and an application to boosting,'' \emph{Journal of Computer and
  System Sciences}, vol.~55, no.~1, pp. 119--139, 1997. [Online]. Available:
  \url{https://www.sciencedirect.com/science/article/pii/S002200009791504X}
\BIBentrySTDinterwordspacing

\bibitem{Friedman}
\BIBentryALTinterwordspacing
J.~H. Friedman, ``Stochastic gradient boosting,'' \emph{Computational
  Statistics \& Data Analysis}, vol.~38, no.~4, pp. 367--378, 2002, nonlinear
  Methods and Data Mining. [Online]. Available:
  \url{https://www.sciencedirect.com/science/article/pii/S0167947301000652}
\BIBentrySTDinterwordspacing

\bibitem{Natekin}
\BIBentryALTinterwordspacing
A.~Natekin and A.~Knoll, ``Gradient boosting machines, a tutorial,''
  \emph{Frontiers in Neurorobotics}, vol.~7, 2013. [Online]. Available:
  \url{https://www.frontiersin.org/article/10.3389/fnbot.2013.00021}
\BIBentrySTDinterwordspacing

\bibitem{Neves}
A.~C. Neves \emph{et~al.}, ``A new approach to damage detection in bridges
  using machine learning,'' in \emph{Experimental Vibration Analysis for Civil
  Structures}.\hskip 1em plus 0.5em minus 0.4em\relax Cham: Springer
  International Publishing, 2018, pp. 73--84.

\bibitem{Basheer}
I.~Basheer, ``Selection of methodology for neural network modeling of
  constitutive hystereses behavior of soils,'' \emph{Computer‐Aided Civil and
  Infrastructure Engineering}, vol.~15, pp. 445 -- 463, 12 2002.

\bibitem{Kaggle}
``Flavours of physics: Finding $\tau \rightarrow \mu\mu\mu $,''
  \url{https://www.kaggle.com/competitions/flavours-of-physics/data}, 2015.

\bibitem{Wikipedia}
M.~Thoma, ``Roc-kurve - die abszisse ist die falsch-positiv-rate und die
  ordinate die richtig,''
  \url{https://upload.wikimedia.org/wikipedia/commons/1/13/Roc_curve.svg},
  2018.

\end{thebibliography}
\vspace{12pt}

\end{document}